\documentclass{raa}            

\usepackage{graphicx,times}             

\begin{document}

   \title{The NGC 7742 star cluster luminosity function: A population analysis revisited}

   \volnopage{Vol.0 (201x) No.0, 000--000}      
   \setcounter{page}{1}          

   \author{Richard de Grijs
      \inst{1,2,3}
   \and Chao Ma
      \inst{1,2}
   }

   \institute{Kavli Institute for Astronomy \& Astrophysics, Peking
     University, Yi He Yuan Lu 5, Hai Dian District, Beijing 100871,
     China; {\it grijs@pku.edu.cn}\\
        \and 
      Department of Astronomy, Peking University, Yi He Yuan Lu 5, Hai
      Dian District, Beijing 100871, China\\
        \and
      International Space Science Institute--Beijing, 1 Nanertiao,
      Zhongguancun, Hai Dian District, Beijing 100190, China\\ }

   \date{Received~~2017 month day; accepted~~2017~~month day}

\abstract{We re-examine the properties of the star cluster population
  in the circumnuclear starburst ring in the face-on spiral galaxy NGC
  7742, whose young cluster mass function has been reported to exhibit
  significant deviations from the canonical power law. We base our
  reassessment on the clusters' luminosities (an observational
  quantity) rather than their masses (a derived quantity), and confirm
  conclusively that the galaxy's starburst-ring clusters---and
  particularly the youngest subsample, $\log(t \mbox{ yr}^{-1}) \le
  7.2$---show evidence of a turnover in the cluster luminosity
  function well above the 90\% completeness limit adopted to ensure
  the reliability of our results. This confirmation emphasizes the
  unique conundrum posed by this unusual cluster population.
   }

   \authorrunning{Richard de Grijs \& Chao Ma}            
   \titlerunning{The NGC 7742 cluster luminosity function revisited}  

   \maketitle

%
%
\section{Rapid star cluster evolution}           
\label{sect:intro}

Until recently, neither theoretical predictions nor prior
observational evidence indicated that the power-law cluster luminosity
functions (CLFs) and the equivalent cluster mass functions (CMFs)
commonly found for very young star cluster systems could transform
into the lognormal distributions characteristic of old globular
clusters on timescales as short as a few $\times 10^7$ yr. Yet,
preliminary evidence is emerging of enhanced cluster disruption on
very short timescales in the complex galactic dynamical environments
of circumnuclear starburst rings (for a brief review, see de Grijs et
al. 2017).

Over the past few decades, significant efforts have been expended to
understand the process of star cluster disruption (e.g., Boutloukos \&
Lamers 2003; Gieles et al. 2005; Lamers et al. 2005; Baumgardt et
al. 2008; Fall et al. 2009; Vesperini et al. 2009). Mengel et
al. (2005) sorted the various disruptive processes acting on star
clusters into a number of different types, as a function of cluster
age. Following an initial period of rapid supernova-driven cluster
expansion and dissolution of up to a few $\times 10^7$ yr, cluster
disruption is predominantly driven by internal two-body relaxation
processes and stellar evolution, at least in the absence of
significant external perturbations (which is, however, unlikely to be
the case in the dense and dynamically complex environments of
starburst rings). In turn, in quiescent environments the initial
power-law CLFs and CMFs transform into the well-established
bell-shaped distributions characteristic of old globular clusters on
timescales of billions of years.

The discovery by de Grijs \& Anders (2012) of a lognormal-like CMF for
star clusters as young as $\sim 10$ Myr in the circumnuclear starburst
ring galaxy NGC 7742 was therefore unexpected. de Grijs \& Anders
(2012) argued that, except if one releases the assumption that the
initial CMF was a power law---which would contradict most
observational studies in this very active area of current
research---three effects could potentially have caused this
discrepancy: (i) technical issues related to the cluster age and mass
determinations, (ii) evolution of the star cluster population on very
short timescales ($\la 10^7$ yr), and/or (iii) differences in the
cluster formation conditions compared with other environments known to
host large samples of young star clusters. They ruled out an origin
related to their analysis approach (i.e., option i) based on a careful
technical assessment.

The community's commonly accepted point of view has gradually evolved
to embrace the idea that star clusters form following a power-law
CMF. This power-law form seems indeed ubiquitous and appears to be
independent of the initial stellar and gas densities. Therefore, an
initial power-law CMF seemed a reasonable, physically motivated
boundary condition for the observed CMF in NGC 7742, and one can thus
rule out option (iii) above (for detailed arguments, see de Grijs \&
Anders 2012). This leaves us with option (ii) as the most likely
origin of the observed peaked CMF. This initial conclusion was
strengthened by the recent study of V\"ais\"anen et al. (2014), who
discovered similar evidence of rapid star cluster disruption in the
early-type galaxy NGC 2328.

\section{Cluster mass versus luminosity functions}

However, we have long been concerned about the high mass at which the
peak of the apparent CMF turnover in NGC 7742 seems to occur, $\log(
M_{\rm cl} / M_\odot) \approx 5.6$. Our concerns were fanned by
similar criticism voiced by other active scientists in the field,
which---as we will see in this paper---was indeed justified to some
extent. We therefore decided to re-analyze the NGC 7742 cluster
population by focusing exclusively on its {\it luminosity} function as
the most straighforward diagnostic tool that does not require one to
adopt the kind of assumptions needed to derive cluster masses. We
proceeded entirely independently from our earlier work (the data
re-reduction was performed indepently as well), to the extent that we
even downloaded the latest pipeline-processed {\sl Hubble Space
  Telescope} ({\sl HST}) images from the Hubble Legacy
Archive.\footnote{http://hla.stsci.edu/hlaview.html} The data
reduction procedures employed followed the same general approach as
those adopted by de Grijs et al. (2013, 2017).

\begin{figure}[h!]
\begin{center}
\includegraphics[width=\columnwidth]{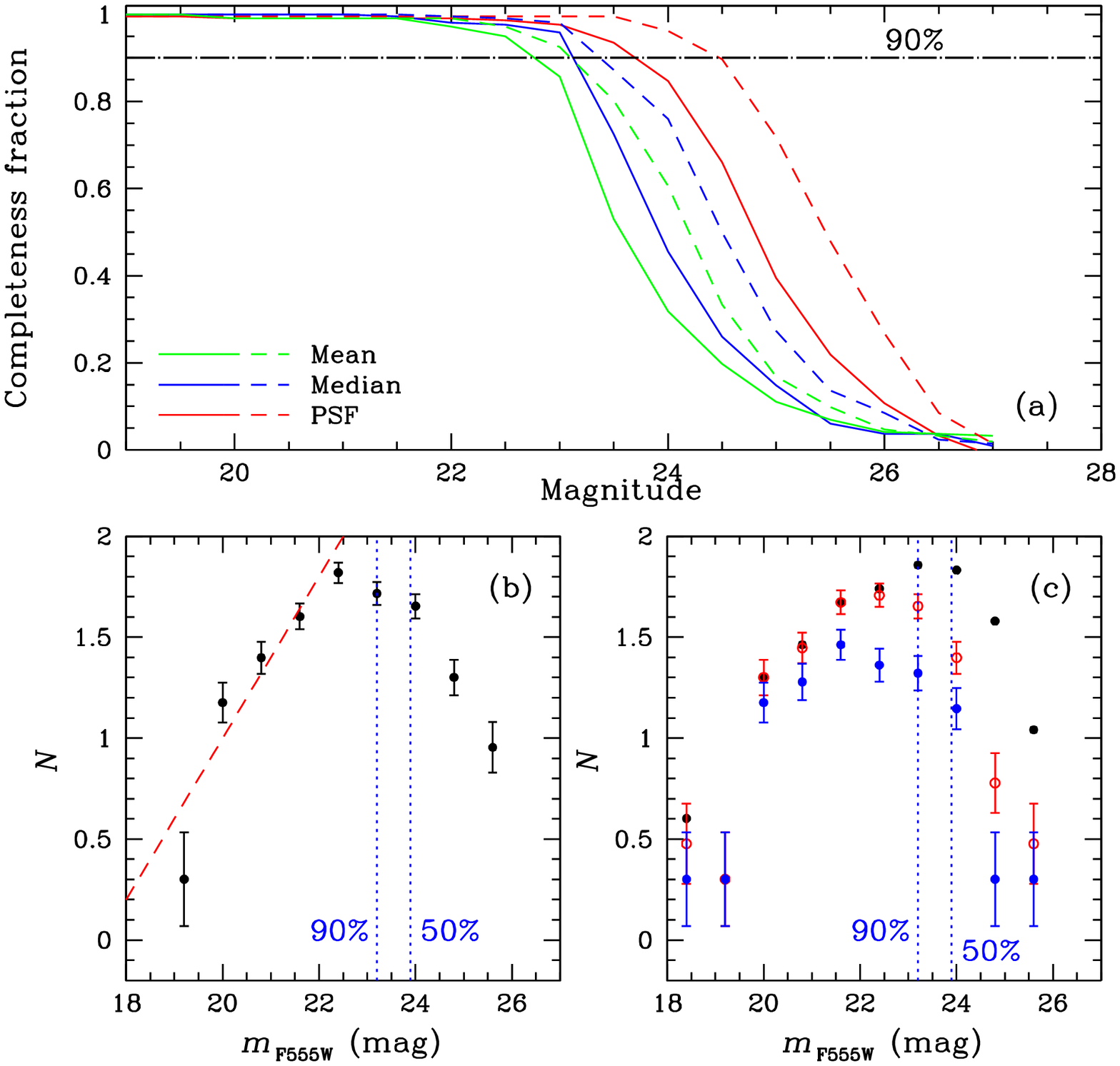}
\caption{NGC 7742 re-analysis. (a) Completeness fractions for point
  sources (`PSF'), median, and mean sizes (FWHM = 1.7, 2.2, and 2.6
  pixels, respectively) in the NGC 7742 starburst ring, at
  galactocentric radii $0.52 \le R \le 1.35$ kpc (corresponding to the
  ring region adopted by de Grijs \& Anders 2012) in the F555W (solid
  lines) and F814W (dashed lines) filters used for the initial source
  selection and spatial cross correlation. (b) CLF of the full sample
  of starburst-ring clusters, based on a complete re-analysis of the
  NGC 7742 data sets. The 50\% and 90\% completeness limits in the
  limiting filter (F555W) for clusters of median size pertaining to
  the starburst ring area are shown as vertical blue dotted lines; the
  canonical $\alpha = 2$ power-law CLF is represented by the red
  dashed line. (c) As panel (b) but for the original data analyzed by
  de Grijs \& Anders (2012), adjusted for the zero-point difference
  derived in our re-analysis (see the text). The black data points
  represent the CLF of the entire cluster population (uncertainties
  are not included for reasons of clarity), while the red CLF relates
  to the ring's subsample. The blue CLF shows the distribution of
  clusters with ages $\log( t \mbox{ yr}^{-1}) \le 7.2$.}
\label{n7742.fig}
\end{center}
\end{figure}

\subsection{Completeness analysis}

In the context of our CLF shape analysis, it is crucial to understand
how many objects may have been missed in a statistical sense by our
processing technique. We employed our previously well-established
approach to determine the extended cluster sample's incompleteness
(e.g., de Grijs et al. 2017; and references therein), of which we
summarize the salient features here. We first applied the {\sc Iraf}
tasks {\sc starlist} and {\sc mkobj} to generate 500 artificial
sources with FWHM = 1.7, 2.6, and 3.4 pixels, corresponding to
approximately 19, 29, and 37 pc, respectively, at the distance of NGC
7742, $D = 22.2$ Mpc. These sizes represent true point sources and
objects with the sample's median and mean sizes,
respectively. Two-thirds of our sample objects have sizes smaller than
the mean size; any larger objects are confirmed cluster complexes. We
chose to use 10 runs of 500 artificial objects each to minimize the
effects of crowding and blending, both between artificial sources and
between artificial and real objects. We checked and confirmed that
using 800 sources or more did not change the final completeness
results. We also checked, specifically for the point-source tests,
that using realistic WFPC2 point-spread functions instead of Gaussian
point sources did not lead to statistically significant changes to the
results. Finally, both authors performed the completeness analysis
independently to cross check the reliability of the results.

We sampled our artificial objects at random $(X, Y)$ coordinates,
ensuring that they were easily identifiable as single sources (that
is, they were not affected by either crowding or blending); 260 of the
500 input sources were located in the ring region defined by de Grijs
\& Anders (2012): $50 \le R \le 125$ pixels. Next, we generated a
blank template image without any background noise and of the same size
as that of the actual NGC 7742 field of view considered here. We then
added the artificial sources located in the ring region to the
template image. All artificial objects were assigned the same
magnitude; we repeated this process by varying the magnitudes of the
input sources from 19.0 mag to 27.0 mag in both the F555W and F814W
filters (roughly equivalent to the Johnson--Cousins $V$ and $I$
broad-band filters), in steps of 0.5 mag.

Finally, we re-`discovered' (and re-measured) the number of artificial
sources in both the template and science images, using the same
approach for both. (We applied our source discovery routine to the
science images to find the real clusters using the exact same
approach.) Since the template image contained no background noise, we
can recover most input artificial sources to within the associated
measurement uncertainties, although some may have disappeared into the
physical noise of the science images, depending on the combinations of
their integrated magnitudes and extent, as well as on the presence of
real background variations. In addition, saturated sources and
blending may cause failures to recover the artificial sources. The
template image hence became a perfect comparison sample to check how
many sources should be recovered under ideal conditions. By
redetermining the photometric and structural properties (i.e., the
Gaussian sizes) of the artificial sources in the combined science and
template images as a function of input magnitude, we get a
quantitative handle on the effects of background noise. We counted the
number of recovered artificial sources with magnitudes and Gaussian
$\sigma$'s within the 1$\sigma$ uncertainties of the input magnitudes
(from the science+template image) and the number of recovered
artificial sources of the same photometric quality and with similar
sizes (from the template) to estimate the completeness, i.e. $f_{\rm
  comp} = N_{\rm rec}/N_{\rm tot}$, where $N_{\rm rec}$ is the number
of artificial objects we recovered and $N_{\rm tot}$ is the number of
input objects.

Our approach ensured that $f_{\rm comp}$ was simultaneously corrected
for the effects of blending and saturation, as well as background
noise. Since our artificial sources are extended objects, correcting
for the effects of crowding and blending is important. We corrected
for and removed blends of artificial sources.

Figure \ref{n7742.fig}a shows the completeness curves in the starburst
ring area for sources with sizes equivalent to the PSF, median, and
mean cluster sizes in the F555W and F814W images, which were initially
used for our selection of genuine sources and their spatial cross
correlation. This panel indicates that the F555W filter is our
limiting passband. In our recent papers (e.g., de Grijs et al. 2017;
and references therein) we have adopted the 90\% completeness level as
our threshold for source detections to underscore the robustness of
our results. These levels occur at $\sim 0.5$--1.0 mag brighter
magnitudes than the conventionally used 50\% completeness limits, in
all filters. The 90\% completeness limit in the NGC 7742 ring area is
reached at F555W magnitudes of $m_{\rm F555W} = 23.7, 23.2$, and 22.8
mag for point sources and objects with the cluster population's median
and mean sizes, respectively.

Our independent approach uncovered 280 cluster candidates associated
with the galaxy's starburst ring, a number that is sufficiently close
to the 256 ring clusters identified by de Grijs \& Anders (2012) to
confirm our confidence in the earlier sample selection. The full ring
CLF is shown in Fig. \ref{n7742.fig}b, where we have also indicated
the 50\% and 90\% completeness levels and the expected, canonical
$\alpha = 2$ power-law CLF for a single-aged cluster population
shortly after cluster formation. We point out that this latter
condition is not met by the full cluster population in the NGC 7742
starburst ring (de Grijs \& Anders 2012; their Fig. 1). We show the
canonical power-law CLF simply for guidance.

\subsection{Direct comparison with earlier results}

A direct comparison of the F555W CLF of Fig. \ref{n7742.fig}b with the
CLF underlying the data presented by de Grijs \& Anders (2012) raised
an immediate concern: while the {\it shapes} of both CLFs were
similar, the {\it luminosities} of the main features differed
significantly, in the sense that the turnover in the newly derived CLF
occurred at a $\sim$2.1 mag fainter level (when compared with the peak
in the CMF, expressed in magnitudes, by adopting a suitable
age-dependent mass-to-light ratio). Careful examination of the
potential cause(s) of this difference revealed that the image header
keywords {\sc photflam} adopted in de Grijs \& Anders (2012) were
incorrectly populated in the original {\sl HST} data frames (STScI
n.d.). We hence adjusted our original photometry following the {\sl
  HST} project team's recommendations, resulting in the CLFs shown in
Fig. \ref{n7742.fig}c.

Figure \ref{n7742.fig}c shows the full NGC 7742 CLF (black), as well
as the CLF for the starburst ring only (red), where we used the same
radial range as de Grijs \& Anders (2012) to define the ring region,
i.e., $0.52 \le R \le 1.35$ kpc. The latter CLF is appropriate for
comparison with the CLF shown in Fig. \ref{n7742.fig}b. Within the
statistical uncertainties, both CLFs are indeed similar. Our cluster
sample exhibits a clear although broad luminosity--size relation (not
shown), in the sense that more extended objects tend to be
brighter. Therefore, even if we somehow missed a sizeable number of
more extended sources will unlikely explain the observed turnover,
given that such extended sources would be missed at the bright end of
the distribution. In view of the available observational evidence, we
therefore conclude that the observed turnover in the CLF is real and
not owing to technical aspects associated with our data reduction
approach.

Unfortunately, our NGC 7742 data set does not include H$\alpha$
images, which would have allowed us to independently confirm the
clusters' young ages, as we did in de Grijs et al. (2017) for a number
of other young starburst ring CLFs. Therefore, we used the cluster
ages derived by de Grijs \& Anders (2012) to make a further selection
of the youngest clusters only. Following de Grijs \& Anders (2012), we
selected all clusters with ages $\log( t \mbox{ yr}^{-1}) \le 7.2$ as
representative of the youngest CLF. The latter is shown as the blue
CLF.

\section{Verdict}

Keeping in mind that our age estimates are based on broad-band
spectral-energy distributions composed of only four filters (which
hence implies that they are affected by significant uncertainties; see
de Grijs \& Anders 2012), it is clear that the 90\% completeness limit
occurs at a brightness level that is at least 1 mag fainter than the
onset of the apparent turnover. It thus appears that while our mass
estimates were rendered too high by the incorrectly populated {\sl
  HST} header keywords, the {\it shape} of the young CLF resembles
that of the young cluster {\it mass} function. Despite our more
thorough analysis presented here, we have thus been unable to
discredit our earlier analysis of the NGC 7742 CMF beyond a reasonable
doubt. The galaxy's young ring CLF and CMF indeed exhibit clear and
robust turnovers that are unexpected from standard cluster evolution
theory. 

We have compiled a catalog of well-resolved nuclear rings that have
been observed with both the {\sl HST} in at least four filters (ideal
for our star cluster analysis) and the {\sl Spitzer Space Telescope}
(see Ma et al. 2017). NGC 7742 is an outlier in our ring
catalog. Traditional nuclear rings are formed as the result of
bar-driven gas inflows, which makes them resemble two winded spirals;
their host galaxies are usually barred spirals (see, e.g., NGC 1512
and NGC 6951; de Grijs et al. 2017) and their CLFs (CMFs) do not
exhibit any obvious turnovers. However, the circumnuclear starburst
ring in NGC 7742 (an unbarred face-on spiral galaxy), although similar
in size to the rings in other ring galaxies, is almost fully circular
and highly symmetric. 

The mass of the circumnuclear ring in NGC 7742, $M_{\rm ring} \sim 6
\times 10^9 M_\odot$, is large with respect to the stellar mass of the
galaxy as a whole, $M_{\rm gal} \sim 5.9 \times 10^{10} M_\odot$
(C. Ma et al., in prep.). The latter value is based on the galaxy's
mass-to-light ratio published by Bell et al. (2003) and its integrated
$K$-band magnitude from Huchra et al. (2012); we determined the ring
mass following the same approach as Ma et al. (2017). An origin
related to a minor merger event has been proposed as an alternative to
the bar-driven gas flows usually invoked (Sil'chenko \& Moiseev 2006;
see also Mazzuca et al. 2006; Tutukov \& Fedorova 2006). Further
investigation of a larger sample of ring galaxies will be required to
shed conclusive light onto the processes driving the establishment of
the unusual type of young CLF (CMF) first uncovered in NGC 7742. A
partial reassessment of our current understanding of the star cluster
evolution scenario may well be on the cards.

\begin{acknowledgements}
This paper is based on observations made with the NASA/ESA {\sl HST},
and obtained from the Hubble Legacy Aarchive, which is a collaboration
between the Space Telescope Science Institute (STScI/NASA), the Space
Telescope European Coordinating Facility (ST-ECF/ESA), and the
Canadian Astronomy Data Centre (CADC/NRC/CSA). This research has also
made use of NASA's Astrophysics Data System Abstract Service. This
work was supported by the National Key Research and Development
Program of China through grant 2017YFA0402702. We also acknowledge
research support from the National Natural Science Foundation of China
(grants U1631102, 11373010, and 11633005).
\end{acknowledgements}

\label{lastpage}


\begin{thebibliography}{99}

\bibitem[2008]{Baum08} Baumgardt H., Kroupa P., Parmentier G., 2008,
  MNRAS, 384, 1231

\bibitem[2003]{Bell03} Bell E.~F., McIntosh D.~H., Katz N., Weinberg
  M.~D., 2003, ApJS, 149, 289

\bibitem[2003]{Bout03} Boutloukos S.~G., Lamers H.~J.~G.~L.~M., 2003,
  MNRAS, 338, 717

\bibitem[2012]{degr12} de Grijs R., Anders P., 2012, ApJL, 758, L22

\bibitem[2013]{degr13} de Grijs R., Anders P., Zackrisson E., \"Ostlin
  G., 2013a, MNRAS, 431, 2917

\bibitem[2017]{degr17} de Grijs R., Ma C., Jia S., Ho L. C., Anders
  P., 2017, MNRAS, 465, 2820

\bibitem[2009]{Fall09} Fall S.~M., Chandar R., Whitmore B.~C., 2009,
  ApJ, 704, 453

\bibitem[2005]{Giel05} Gieles M., Bastian N., Lamers H.~J.~G.~L.~M.,
  Mout J.~N., 2005, A\&A, 441, 949

\bibitem[2012]{Huch12} Huchra J.~P., Macri L. M., Masters K. L., et
  al., 2012, ApJS, 199, 26

\bibitem[2005]{Lame05} Lamers H.~J.~G.~L.~M., Gieles M., Portegies
  Zwart S.~F., 2005, A\&A, 429, 173

\bibitem[2017]{Ma2017} Ma C., de Grijs R., Ho L. C., 2017, ApJS, 230,
  14

\bibitem[2006]{Mazz06} Mazzuca L.~M., Sarzi M., Knapen J.~H., Veilleux
  S., Swaters R., 2006, ApJL, 649, L79

\bibitem[2005]{Meng05} Mengel S., Lehnert M.~D., Thatte N., Genzel R.,
  2005, A\&A, 443, 41

\bibitem[2006]{silc06} Sil'chenko O.~K., Moiseev A.~V., 2006, AJ, 131,
  1336

\bibitem[n.d.]{STScI} STScI, n.d., WFPC2 Advisory: Incorrect PHOTFLAM
  header keyword value in Multidrizzle output
  images. http://www.stsci.edu/hst/wfpc2/documents/photflam.pdf

\bibitem[2006]{Tutu06} Tutukov A.~V., Fedorova A.~V., 2006,
  Astron. Rep., 50, 785

\bibitem[2014]{Vais14} V{\"a}is{\"a}nen P., Barway S., Randriamanakoto
  Z., 2014, ApJL, 797, L16

\bibitem[2009]{Vesp09} Vesperini E., McMillan S.~L.~W., Portegies
  Zwart S., 2009, ApJ, 698, 615

\end{thebibliography}
\end{document}